\documentclass[twocolumn,aps,showpacs]{revtex4}
\usepackage{graphicx}

\begin{document}
\title{Theory of radiation trapping by the accelerating solitons in optical fibers}

\author{Andrey V. Gorbach}
\affiliation{Centre for Photonics and Photonic Materials, Department
of Physics, University of Bath, Bath BA2 7AY, UK}
\author{Dmitry V. Skryabin}
\affiliation{Centre for Photonics and Photonic Materials, Department of Physics,
University of Bath, Bath BA2 7AY, UK}

\begin{abstract}
We present a theory describing trapping of the normally dispersive
radiation by the Raman solitons in optical fibers. Frequency of the
radiation component is continuously blue shifting, while the soliton
is red shifting. Underlying physics of the trapping  effect is in
the existence of the inertial gravity-like force acting on light in
the accelerating frame of reference.  We present analytical
calculations of the rate of the opposing frequency shifts of the
soliton and trapped radiation  and find it to be greater than the
rate of the red shift of the bare Raman soliton. Our findings are
essential for understanding of the continuous shift of the high
frequency edge of the supercontinuum spectra generated in photonic
crystal fibers towards higher frequencies.
\end{abstract}


\maketitle

\section{Introduction}
Frequency conversion in optical fibers has been an active research field
already for few decades \cite{agrawal}. Most striking and extensively studied
recent advance has been generation of extremely broad optical spectra (supercontinua)
in optical fibers with small effective area,
pumped by femto-second pulses with the carrier frequency
close to the point of the zero group velocity dispersion (GVD) \cite{RWS2000, DGX+2002}.
Applications of the supercontinuum include spectroscopy, metrology \cite{HUH+2000},
telecommunication \cite{SAE+2006} and medicine \cite{HLC+2001}.

Amongst problems posed by the observation of supercontinuum one of
the most puzzling has been  understanding of the nonlinear processes
leading to the generation of the high frequency wing of the
supercontinuum continuously drifting towards even higher frequencies
\cite{HH2001, WRW2002, GLL+2002, TCD2003, ASE+2006}. Several
experimental and numerical observations  explicitly demonstrated
that the radiation at the blue wing of the supercontinuum often
propagates in the form of nondispersive wave packets, localized on
the femtosecond scale and continuously blue shifting \cite{GLL2004,
HNG+2004, GSS+2006, FFB2005}. Note that GVD at the blue edge of the
continuum is typically normal, therefore the dispersive spreading
can not be compensated by the nonlinearity. Independently from the
supercontinuum generation the effect of the localization of blue shifting pulses
in the normal GVD range, coupled to the Raman solitons propagating in the
anomalous GVD range, has been reported in the series of papers by
Nishizawa and Goto \cite{NG2002b, NG2002, NG2002a} and more recently
by Cheng and co-authors \cite{CWF+2005}. It has been proposed in
Refs. \cite{NG2002b, NG2002, NG2002a, CWF+2005} that the physical
mechanism behind the above effects is the cross-phase modulation
(XPM) \cite{agrawal,Manassah1988}.

However, it is well known that the XPM coupling between anomalously
and normally dispersing  components can lead to dispersion
compensation and formation of the bright-dark soliton pairs  only if
one soliton component is a dark pulse and the other one is bright,
see, e.g., \cite{TWW+1988, Kivshar1992, BKP1996}. Thus the XPM  can
not be the sole reason for formation of the bright-bright localized
states across the zero GVD wavelength \cite{GLL2004, HNG+2004,
GSS+2006,FFB2005,NG2002b,NG2002, NG2002a}. Also,  the red shift of
the anomalously dispersing component is readily explained by the
intrapulse Raman scattering \cite{agrawal,Gordon1986}, while the
blue shift of the normally dispersing bright pulse coupled to it
requires to be understood.

Reference  \cite{GSS+2006}  has explained formation  of the blue edge
of  supercontinua in fibers  using the  theory of four-wave mixing
between the solitons and dispersive waves  \cite{SY2005, YSR2004}.
It has been demonstrated that for  typical fiber dispersions
the interaction between the soliton and the blue radiation happens recurrently, so that every
scattering event leads to the further blue shift of the signal
pulse \cite{GSS+2006}. Though this theory describes well first stages of the blue edge formation
it fails to explain why the femtosecond pulses emerging there remain free of dispersive spreading.
The latter is naturally expected because of
the strong normal GVD and would lead to a fast degradation  of any
nonlinear interaction between the soliton and the blue radiation,
which in practice continues over long distances.
In our recent work \cite{GS2007} we have explained the physical
mechanisms behind existence of the non-dispersive and continuously
blue shifting localized states of light on the high frequency edge
of the supercontinuum spectra. The light is trapped by the
refractive index changes induced, on one side (front of the pulse),
by the red shifting Raman solitons via the nonlinear cross coupling and, on the
other side (trailing tail), by the inertial force originating from
the fact that the solitons move with acceleration. The nature of the
latter effect is analogous to the gravity-like inertial force acting
on an observer in a rocket moving with a constant acceleration.

The aim of this work is not only to provide mathematical details for the
mostly qualitative description presented in \cite{GS2007}, but also
to extend the theory into the regime of sufficiently strong
intensities of the blue radiation. In this regime the trapped blue
component of the two frequency bound-state starts to
influence  the  soliton dynamics  on the red edge, which
makes noticeable quantitative impact on the propagation dynamics
of the bound state.

\section{The model}

\begin{figure}
\includegraphics[width=0.48\textwidth]{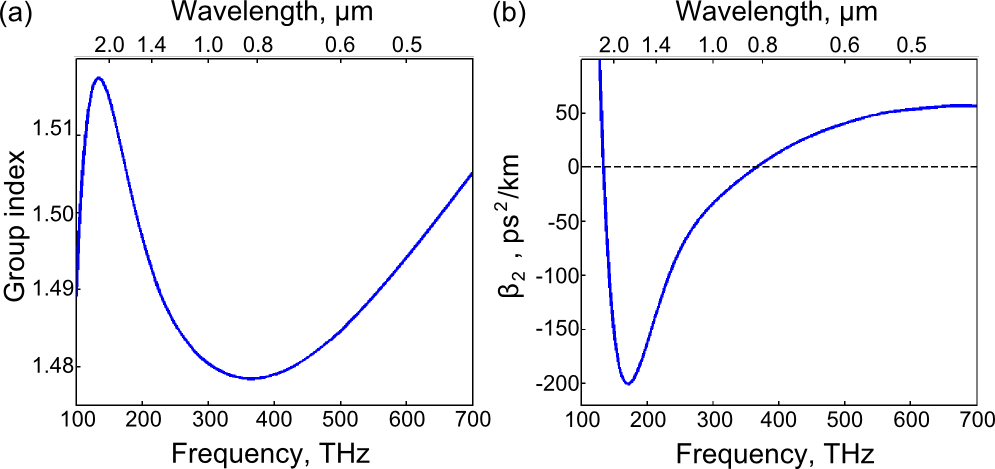}
\caption{(Color online)
Frequency dependence of the group velocity normalized to speed of light in vacuum (group index)
(a) and of the GVD (b) in the photonic crystal fiber as in Ref. \cite{GSS+2006}.
Fiber dispersion like this is typical for numerous supercontinuum experiments.}
\label{fig1}
\end{figure}

The subject of this work is the  explanation of the existence and
detailed study of the previously reported \cite{GLL2004, HNG+2004,
GSS+2006, FFB2005, NG2002b, NG2002, NG2002a, CWF+2005} two-frequency
bright-bright soliton-like states in optical fibers, with frequency
of one component being in the anomalous and frequency of the other
being in the normal GVD range. Thus the dispersion we need to use
should include the sign change of the GVD. The simplest example of
the dispersion operator having this property is
\begin{equation}
\partial_zE-{1\over 6}\epsilon\partial_t^3E=0\;.
\label{cubic_disp}
\end{equation}
Assuming $E=e^{ikz-i\delta t}$, we find that the GVD is
$\partial_{\delta}^2k=\epsilon\delta$. Thus $\delta$ measures the
spectral deviation from the zero GVD point at $\delta=0$. The
third-order dispersion coefficient $\epsilon$ is positive for the
telecom fibers and in the proximity of the $800$nm for the typical
photonic crystal fiber (PCF) designs used for generation of
supercontinuum with femtosecond pulses \cite{RWS2000,DGX+2002}.
$\epsilon$ can also be negative in the proximity of $1500$nm in PCFs
and tapered fibers with sufficiently small cores \cite{SLK+2003}. It
is important to note  that the group index parameter
$\partial_{\delta}k=\epsilon\delta^2/2$ is symmetric with respect to
$\delta\to-\delta$ and therefore the group velocity is matched
across the zero GVD point. The matching, or near matching, of the
group velocities is important for the existence of the two-frequency
bound states, because the pulses should not spatially separate
before the bound state is established. In the
PCFs the third order dispersion $\partial^3_{\delta}k$ usually
changes its value and sign as frequency varies, as can be seen from
the change of slope of the GVD curve in Fig.~\ref{fig1}(b). This
eventually destroys the matching of the group velocities across the
zero GVD point, see Fig.~\ref{fig1}(a). However, the fact that the
matching is still achieved over the wide bandwidth is the most
important for the effect of radiation trapping and for the formation
of the blue wing of a supercontinuum.

Taking into account the instantaneous Kerr nonlinearity and  the
non-instantaneous Raman response, the light
propagation in a fiber is modeled by the dimensionless generalized
nonlinear Schr\"odinger (NLS) equation \cite{agrawal}
\begin{eqnarray}
\nonumber &&\partial_{z}E=i\hat k(i\partial_t)E+
i(1-\theta)E|E|^2\\
&&+E\int_{-\infty}^{+\infty}{R(t')|E(t-t',z)|^{2}}dt'.\label{eqn1}
\end{eqnarray}
The dispersion operator in Eq. (\ref{eqn1}) is
\begin{equation}
\hat k(i\partial_{t})\equiv \sum_{m=
3}^M  {L \beta^{(m)}
\over \tau^mm!} [i\partial_{t}]^m, \label{eq_disp}
 \end{equation}
where  the  coefficients $\beta^{(m)}$
are selected to fit the dispersion profile of the fibers. The total electric
field is given by $Ee^{ik_rz-i\omega_r t}+c.c.$, where the reference frequency
$\omega_r$ is chosen to coincide with the zero GVD frequency, that is why the sum in
Eq. (\ref{eq_disp}) starts from $m=3$.
$R(t)$ is the standard Raman response function of silica:
\begin{equation}
R(t)=
\theta{\tau_1^2+\tau_2^2\over\tau_1\tau_2^2}\Theta(t)e^{-t/\tau_2}\sin{t\over\tau_1}\;.
\label{eqn2}\end{equation}
Here  $\Theta(t)$ is the Heaviside
function and parameter $\theta=0.18$ weights the Raman nonlinearity relative to the Kerr one.
Characteristic times of the  delayed Raman response  are
$\tau_1=12.2fs/\tau$ and $\tau_2=32fs/\tau$ \cite{agrawal}.
$t$ is the dimensionless time  in the reference
frame moving with the light group velocity at $\omega_0$
and measured in the units of
$\tau$. $z$ is  the distance along the fiber measured in the units of
$L$, where $L$ is any convenient characteristic length.
Field amplitude  $E$ is  scaled to
$1/\sqrt{\gamma L}$, where $\gamma$ is the nonlinear
parameter of the fiber \cite{agrawal}.
To get feel for the real values of the parameters we choose $\gamma=0.02$(Wm)$^{-1}$ and $L=8$m,
which gives $1/(\gamma L)= 6.25$W.
Choosing $\tau=200$fs and  $\beta^{(3)}=0.12 ps^3/km$
[corresponding to the dispersion slope at the zero GVD point
close to $800$nm in Fig.~\ref{fig1}(b)] we have $\epsilon=L \beta^{(3)}/\tau^3= 0.12$.

\section{Numerical experiments illustrating
radiation trapping by accelerating solitons}
We proceed describing two sets of numerical experiments.
First, is when supercontinuum evolves from a single pump pulse.
Second, is when the two-frequency bound state is excited
directly by the two pulses.
In the first set we used the realistic fiber dispersion, see Fig.~\ref{fig1}.
However,  to explain the effect of radiation trapping by solitons across the zero GVD point
it is sufficient to consider the simple cubic dispersion as in Eq.~(\ref{cubic_disp}), which
simplifies comparison of analytical and numerical results and is used
throughout the rest of the paper.

\begin{figure}
\includegraphics[width=0.48\textwidth]{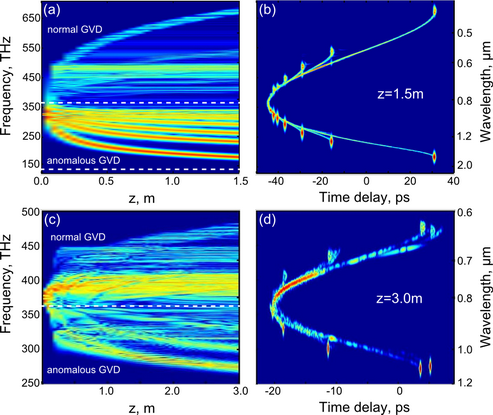}
\caption{
(Color online)
Left panel: spectral evolution along the fiber length for the dispersion profile
as in Fig.~\ref{fig1}. Pump pulse parameters: (a) wavelength $900$nm is in the anomalous GVD range,
 (c) $800$nm is in the normal GVD range. In both cases peak power is $6$kW, duration $200$fs;
 The radiation at the short wavelength edge of the continuum
 starts its continuous drift towards shorter frequencies at $z>0.2m$ for (a)
 and $z>1m$ for (c). Right panel:
 spectrograms showing simultaneous frequency and time domain images
of the supercontinua in (a) and (c) for $z=1.5m$ and $z=3m$, respectively.
The function plotted has been calculated using the XFROG integral
$I(t,\omega)=\int_{-\infty}^{\infty}$ $dt^{\prime}E_{ref}(t^{\prime}-t)E(t^{\prime})e^{-i\omega t^{\prime}}$.
Here $E_{ref}$ is the $2$ps gaussian pulse. See Ref. \cite{XFROGbook} on more details about the
XFROG (cross-correlation frequency resolved optical gating).
}
\label{fig2}
\end{figure}

Figs.~\ref{fig2} illustrate supercontinuum generation
with a single pump
in the fiber as in Fig.~\ref{fig1}.
The pump frequency was chosen
either in the range of anomalous GVD, Figs.~\ref{fig2}(a),(b), or in the range of normal GVD Figs.~\ref{fig2}(c),(d).
In both cases one can see that, after the initial stages of the
supercontinuum development (see e.g. \cite{GSS+2006}, for detailed
description),  the blue tip of the spectrum starts its continuous
drift towards higher frequencies, see $z>0.2m$ in
Fig.~\ref{fig2}(a) and
$z>1m$ in Fig.~\ref{fig2}(c). This spectral shift appears to be
correlated with the soliton self-frequency shift  at the
opposite (infrared) edge of the spectrum. Spectrograms showing the signals simultaneously
in the frequency and time domains, see Figs.~\ref{fig2}(b) and (d), unambiguously demonstrate
that the high frequency tip of the continuum is localized in the time domain
on the same femtosecond scale as the soliton at the infrared edge.
One can also see, that not only the soliton at the very
edge of the spectrum, but all the red shifting solitons have
associated localized pulses on the high frequency side of the spectrum in
the normal GVD range.

\begin{figure}
\includegraphics[width=0.48\textwidth]{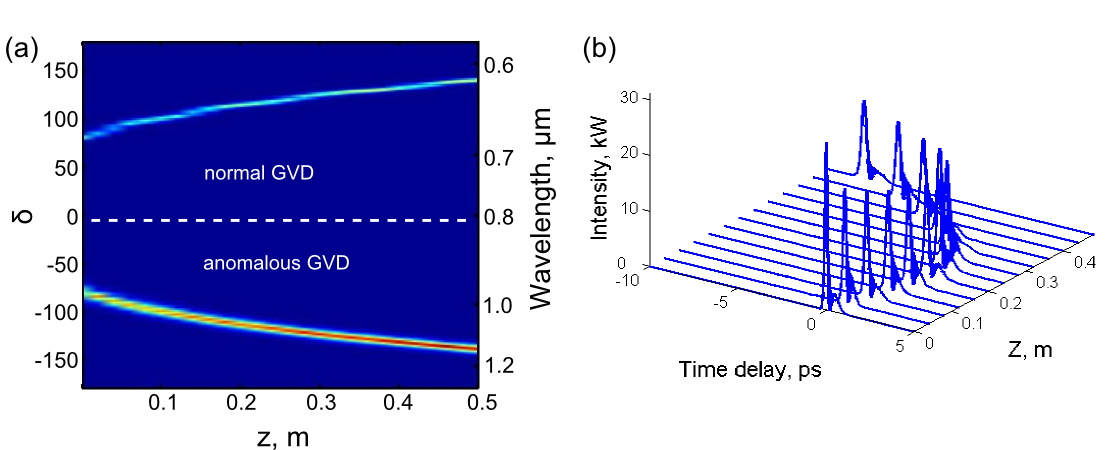}
\caption{
(Color online)
Spectral (a) and time-domain (b) evolution along the fiber pumped
with two pulses obtained from the numerical modeling of Eq. (\ref{eqn1})
with $\beta^{(m>3)}=0$ and $\epsilon=+0.12$. Initial conditions are
$E(z=0,t)=\sqrt{2q_1}\textrm{sech}(t/\tau_1)\exp(-i\delta_1 t)+
\sqrt{2q_2}\textrm{sech}((t-T_0)/\tau_2)\exp(-i\delta_2 t)$.
with $q_1= 2000$ (peak power $\sim 24 kW$), $q_2= 250$
(peak power $\sim 3 kW$), $\tau_1=\sqrt{-D^{\prime\prime}_s/2q_1}=0.2$ ($\sim 40 fs$),
$\tau_2=\sqrt{-D^{\prime\prime}_s/2q_2}\approx 0.57$
($\sim 115 fs$), $T_0=0.25$ ($\sim 50 fs$) and $\delta_1=-80$ ($\sim 964 nm$), $\delta_2=80$
($\sim 684 nm$). Results in (b) are presented in the accelerating frame of reference: $t^{\prime}=t-g_0z^2/2$,
with  $g_0$ defined by Eq.~(\ref{g0}).}
\label{fig3}
\end{figure}

To isolate the effect of formation of the two-frequency bound-states across
the zero GVD point we perform simulations using the two pulse
excitation. From now on we will focus on the simple case of cubic dispersion as in Eq.~(\ref{cubic_disp}).
We start from the case $\epsilon>0$, see Fig.~\ref{fig3}.
The first of the pump pulses is spectrally located in the anomalous GVD regime
and  forms a red shifting Raman soliton. The second pulse is delayed with respect
to the first one and has spectrum in the range of normal GVD.
In the course of the propagation the latter pulse appears
to be trapped on the trailing tail of the soliton, while its frequency
is continuously increasing, see Fig.~\ref{fig3}.
Note that the Raman effect pulls the soliton toward smaller frequencies and
away from the zero GVD point,
therefore the GVD felt by the soliton is continuously increasing. This leads to the
noticeable temporal broadening of the soliton and to the drop in its amplitude,
which also affects the shape of the trapped radiation, see
Fig.~\ref{fig3}(b).
Clearly the group velocities of the two components within the bound state
are the same, which suggests that the frequencies are changing along the dispersion curve
in such a way that the group velocity matching is preserved.

\begin{figure}
\includegraphics[width=0.48\textwidth]{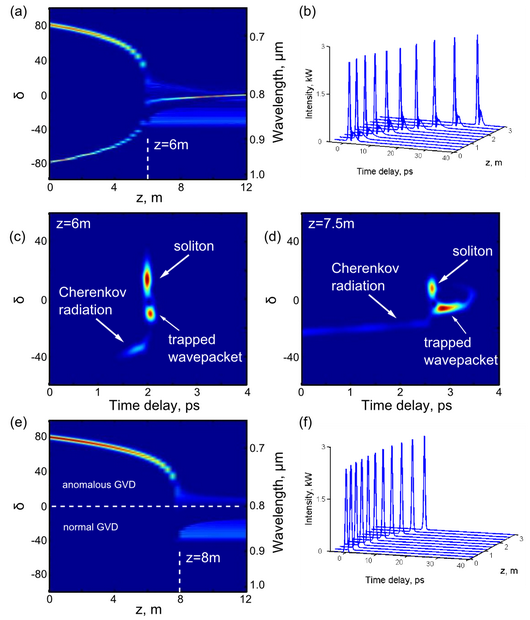}
\caption{
(Color online)
Pulse propagation in the case of negative third-order dispersion,
$\epsilon=-0.12$.
(a) and (b): two pulses pump similar to that in Fig.~\ref{fig3}.
Parameters of the pulses are: $q_1= 200$ (peak power $=\sim 2.4 kW$), $q_2= 25$
(peak power $\sim 0.3 kW$), $\tau_1=\sqrt{-D^{\prime\prime}_s/2q_1}\approx 0.63$
($\sim 130 fs$),
$\tau_2=\sqrt{-D^{\prime\prime}_s/2q_2}\approx 1.8$
($\sim 360 fs$), $T_0=1.5$ ($\sim 300 fs$) and $\delta_1=80$, $\delta_2=-80$.
(c) and (d): XFROG spectrograms for the case of two-pulse pump, calculated
at different propagation distances.
(e) and (f): soliton propagation with $q=200$ (peak power $\sim 2.4 kW$).
All time domain results are presented in the accelerating frame of reference: $t^{\prime}=t-g_0z^2/2$,
with  $g_0$ defined by Eq.~(\ref{g0}).}
\label{fig4}
\end{figure}

In the case of the negative third-order
dispersion ($\epsilon<0$) the effect of the radiation trapping is also clearly
observed, see Fig.~\ref{fig4}.
The difference here is that the normally and anomalously dispersing
components of the two-frequency bound-state are now spectrally
converging  towards the zero GVD point, see Fig.~\ref{fig4}(a).
However, the direction of the spectral evolution of each of the
components  is the same as for $\epsilon>0$.
Namely the Raman effect pulls the soliton component in the anomalous GVD range
towards smaller frequencies, while the pulse in the normal GVD range
shifts towards higher frequencies. The GVD felt by the soliton is reducing in this case,
therefore the soliton is adiabatically compressed, see Figs.~\ref{fig4}(b), (f).
When the soliton is pulled sufficiently close
to the zero GVD point, so that the significant part of its own spectrum is in the
normal GVD regime, it starts to emit Cherenkov radiation \cite{WCL1990, Karpman1993, AK1995},
see Figs.~\ref{fig4}(a), (e).
This radiation creates spectral recoil effect, which counterbalances
the Raman self-frequency shift and stabilizes the soliton frequency \cite{SLK+2003, BSY2004}.
As the soliton looses its energy through radiation, it broadens and so does the trapped wavepacket.
Still, both spectral components of the bright-bright quasi-soliton pair remain fairly well localized over
propagation distances much larger than the GVD length. Spectrograms showing the effect
of the radiation trapping in this case are shown in Figs.~\ref{fig4} (c), (d).
 Note, that the stabilization of the frequency of the pure soliton, i.e.
without the trapped radiation, happens at longer propagation distances,
cf. Figs.~\ref{fig4}(a) and (e). This suggests  that the trapping effect boosts the rate of the
soliton self frequency shift. We will discuss this more in details in Sec. VI.

\section{Coupled NLS equations in the accelerated frame of reference}
In order to explain the above numerical observations
of the two-frequency bound states across the zero GVD point,
we reduce Eq. (\ref{eqn1}) to the coupled NLS
equations for the two pulses on the opposite
sides of the zero GVD frequency. We assume
\begin{equation}
\label{two_waves}
E=A_1\exp[ik_1z-i\delta_1 t]+A_2 \exp [ik_2z-i\delta_2 t],
\end{equation}
where $A_{1,2}$ are the amplitudes of the two pulses, $\delta_{1,2}$ are their frequencies
and $k_{1,2}=k(\delta_{1,2})$ are the wavenumbers: $k(\delta)=e^{i\delta t}\hat k(i\partial_t)e^{-i\delta t}$.
We also assume that $\delta_{1,2}$ are
selected in such a way that $k_1^{\prime}=k^{\prime}_2$ ($k^{\prime}=
\partial_{\delta}k$)
and therefore the group velocities of the two pulses are equal.
Substituting Eq. (\ref{two_waves}) into Eq. (\ref{eqn1}), expanding
$|E(t-t^{\prime})|^2$ in Eq. (\ref{eqn1}) up to the first order Taylor
term and neglecting all the fast oscillating exponential
terms with frequencies $\delta_{1,2}$ and their harmonics we  obtain
a pair of the coupled NLS equations:
\begin{eqnarray}
\label{CNLS1}
\nonumber  i\partial_z A_1+d_1 \partial^2_x A_1&=&-\left[|A_1|^2+2|A_2|^2\right]A_1+\\
& &T A_1 \partial_x \left[|A_1|^2+|A_2|^2\right]\;,\\
\nonumber  i\partial_z A_2+d_2 \partial^2_x A_2&=&-\left[|A_2|^2+2|A_1|^2\right]A_2+\\
& &T A_2 \partial_x \left[|A_1|^2+|A_2|^2\right]\;,
\label{CNLS2}
\end{eqnarray}
where $x=(t-zk_1^{\prime})/\sqrt{|k_1^{\prime\prime}|}$,
$T=|k^{\prime\prime}_1|^{-1/2}\tau_d$ is the effective Raman time,
$\tau_d = \int  tR(t)dt\approx 0.0073$,
$d_1=1/2$ (anomalous GVD), and $d_2=-|k^{\prime\prime}_2|/|2k^{\prime\prime}_1|<0$ (normal GVD).
The most important restriction of Eqs. (\ref{CNLS1}), (\ref{CNLS2}) relative to Eq. (\ref{eqn1})
is that the former do not include frequency dependence of the GVD, see Fig.~\ref{fig6}.
However, this is not critical for  understanding of the trapping mechanisms.

If $A_2= 0$ then Eq.~(\ref{CNLS1})
has an approximate solution in the form of the  NLS soliton
moving with constant acceleration $g$ \cite{agrawal}. So that its center in the $(x,z)$-plane
follows the parabolic trajectory $x=gz^2/2$ and its frequency is continuously
red shifting with the rate $g/(2d_1)$. For the single NLS equation
it has been demonstrated that there exist a transformation
into the accelerating frame of reference, which retains the
structure of the NLS equation apart from  adding a linear in $x$
potential  \cite{CL1976}. $g$ is a free parameter of this symmetry transformation,
which value is fixed by assuming  that the corrections to  the soliton
induced by the linear potential are  balanced by the corrections due to the Raman term, see below.

Here we apply the analogous symmetry transformation to the coupled equations
\begin{eqnarray}
\label{transform_sol}
A_1=\psi\left(z,\xi\right)\exp\left[i\frac{\xi \dot{x}_0}{2d_1}+if_1(z)+i q z\right]\;,\\
\label{transform_rad}
A_2=\phi\left(z,\xi\right)\exp\left[i\frac{\xi \dot{x}_0}{2d_2}+if_2(z)+i \lambda z\right]\;,
\end{eqnarray}
where
\begin{equation}
\label{transform2}
\xi=x-x_0(z),~x_0=gz^2/2,~ f_{1,2}(z)=-\frac{1}{4d_{1,2}}\int \dot{x}_0^2 dz.
\end{equation}
Parameters $q$ and $\lambda$ are the shifts of the propagation constants of the two components.

The resulting equations for $\psi$ and $\phi$ are
\begin{eqnarray}
\label{CNLS1_w}
\nonumber
i\partial_z \psi + d_1 \partial^2_\xi \psi &=&
q \psi - \left[|\psi|^2+2|\phi|^2\right]\psi+\\
& &T \psi \partial_\xi \left[|\psi|^2+|\phi|^2\right] + \frac{g\xi}{2d_1}\psi\;,\\
\label{CNLS2_w}
\nonumber  i\partial_z \phi + d_2 \partial^2_\xi \phi &=&
\lambda \phi - \left[|\phi|^2+2|\psi|^2\right]\phi+\\
& &T \phi \partial_\xi \left[|\psi|^2+|\phi|^2\right] + \frac{g\xi}{2d_2}\phi\;.
\end{eqnarray}
The prime difference of the coupled NLS equations (\ref{CNLS1_w}), (\ref{CNLS2_w})
with the text book ones \cite{agrawal} is the presence of the linear in $\xi$
potentials.  The acceleration $g$ has been the free
parameter upto now. Its selection will be discussed in the next two sections.

We should note here, that the natural next step in our analysis could be the
setting up a boundary value problem for the $z$-independent version of Eqs.
(\ref{CNLS1_w}), (\ref{CNLS2_w}). However, proper setting of the boundary
conditions is a challenging problem due to presence of the linear potential.
The main reason for this is that working out asymptotic behavior
of solutions for $\xi \to -\infty$  requires a  delicate
analysis going beyond the scope of this work. Indeed,
if one simply neglects the nonlinearity, then the tails of both
components behave like Airy functions,
which was assumed in the prior works on the similar problems
\cite{AKL1996, AKV2000, FP2003}.
However, the amplitude of the oscillatory tail ($\xi \to -\infty$)
of the Airy function decays only as $|\xi|^{-1/4}$, which makes
some linear terms in Eqs. (\ref{CNLS1_w}), (\ref{CNLS2_w})
to decay at the rates matching the decay rate of the nonlinear terms,
suggesting that equating the solution tails to the  Airy function
is an  approximation.
This problem still waits for its proper analysis even in the case
of the single soliton equation.
For the above reasons we rely in what follows on perturbation theory
complemented by the numerical calculations with zero boundary conditions for large
$|\xi|$, which  correctly describes the localized parts of the solutions,
but gives only qualitative answers at the oscillatory tails.
The effects we study below are, however, dependent most strongly on the localized
part of the solutions, and therefore our approximation is adequate in the given context.

\section{Linear theory of the radiation trapping}
We proceed, by considering the limit when  the normally
dispersive component $\phi$ is much weaker than the
soliton component $\psi$. In this limit we can
assume that the field $\psi$ is not affected by $\phi$ and
neglect all the terms nonlinear in $\phi$.
Then Eq.~(\ref{CNLS1_w}) for $\psi$ can be solved perturbatively.
We assume $\psi=\psi_0(\xi)+\psi_1(\xi)+\dots$,
where
\begin{equation}
\label{soliton}
\psi_0={\sqrt{2q}}{\textrm{sech}(\sqrt{q/d_1}\xi)}
\end{equation}
 solves $d_1\partial_{\xi}^2\psi_0=q\psi_0-\psi_0^3$ and
$\psi_1$ accounts for corrections due to Raman effect and the linear potential:
\begin{equation}
d_1\partial_{\xi}^2\psi_1-q\psi_1+3\psi_0^2\psi_1
=T\psi_0\partial_{\xi}\psi_0^2+{g\xi\over 2d_1}\psi_0.
\label{psi1}
\end{equation}
The operator $d_1\partial_{\xi}^2-q+3\psi_0^2$ is self-adjoint  and singular.
Its null space is spanned by the single eigenfunction $\partial_{\xi}\psi_0$.
Projecting the right-hand side of Eq. (\ref{psi1}) on the latter we
find
\begin{equation}
g=g_0\equiv {32Tq^2\over 15}\;.
\label{g0}
\end{equation}
Thus the linear potential indeed compensates for the Raman shift,
at least in the first order, providing $g=g_0$.
Note, that this approach neglects the non-exponential
decay of  the soliton tail at $\xi\to-\infty$, see discussion at the end of the
previous section, and it is equivalent to the traditional considerations \cite{Gordon1986,CL1976},
where the oscillatory tail has been ignored.

\begin{figure}
\includegraphics[width=0.48\textwidth]{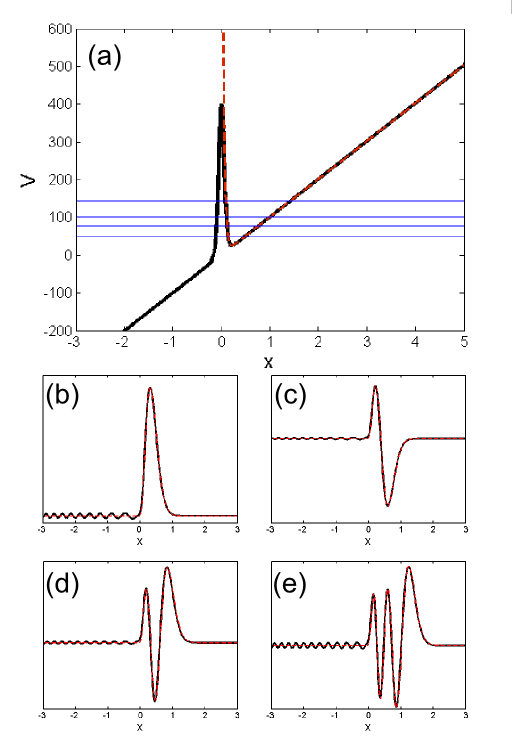}
\caption{
(Color online)
(a) Full (dashed) line shows the potential $V$ ($V_{b}$) for the soliton with $q=100$ and $T=0.0024$.
(b-e) show selected quasi-bound eigenstates  of the potential $V$, cf. horizontal levels in (a).
Dashed lines show the corresponding modes of the potential $V_{b}$.}
\label{fig5}
\end{figure}

The $z$-independent equation for $\phi$
is simply a linear eigenvalue problem in our approximation
\begin{equation}
\label{linear}
-|d_2|\partial^2_\xi\phi + V(\xi) \phi=\lambda\phi,
\end{equation}
where  potential $V(\xi)$ consists of the localized soliton part and of the
linear potential induced by the acceleration with already determined value:
\begin{equation}
\label{linear_pot}
V(\xi)=2\psi_0^2-T \partial_\xi \psi_0^2+\frac{g_0\xi}{2|d_2|}.
\end{equation}
The superposition of the exponentially decaying soliton tail and
of the linear potential creates a local minimum of $V$ on the trailing
tail of the soliton, see Fig.~\ref{fig5}(a),
giving rise to the effect of light localization in the normal GVD regime.
The soliton itself serves as a potential barrier for the normally dispersive waves on one side
of the well and the linear potential, obtained as a result
of the transformation to the accelerated frame
of reference, serves as a barrier on the other side. Thus the accelerating potential
creates an inertial force acting on photons. It is analogous
to the gravity-like force  acting on massive bodies
in a closed container moving with a constant acceleration.
The soliton created potential wall is not infinite, however.
Therefore the light is expected to tunnel through it.
Hence, only quasi-bound states embedded inside the continuum
are possible as solutions of Eq. (\ref{linear}).

\begin{figure}
\includegraphics[width=0.48\textwidth]{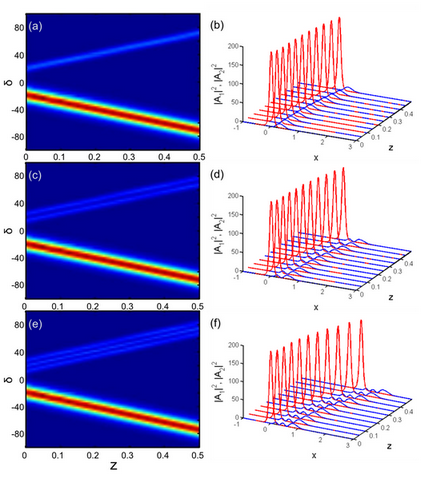}
\caption{
(Color online)
Numerical propagation of the first three bound states within
coupled equations (\ref{CNLS1}), (\ref{CNLS2}).
Initial conditions are: soliton for $A_1$ with $q=100$ and
linear eigenstate of $V$ for $A_2$. Left column:
spectrum of the total field $E$, Eq.~(\ref{two_waves}), with
$\delta_2=-\delta_1=20$. Right column: temporal profiles of $|A_1|^2$ (dashed lines)
and $|A_2|^2$ (solid lines). Time domain results
are presented in the accelerating frame of reference: $t^{\prime}=t-g_0z^2/2$,
with  $g_0$ defined by Eq.~(\ref{g0}).
}
\label{fig6}
\end{figure}

In order to locate these states we first
replace the soliton part of $V$ by its asymptotic and find that
\begin{eqnarray}
\label{approx_linear_pot}
&& V\approx V_{b}\equiv A\exp(-B\xi)+C\xi,\\
\nonumber \label{ABC}
&& A=16q,~ B=2\sqrt{\frac{q}{d_1}},~ C=\frac{g_0}{2|d_2|}.
\end{eqnarray}
Contribution to $A$ due to the Raman term ($\sim T$) is small as compared to $16q$ and is neglected.
The potential $V_{b}$  goes to infinity on both sides and
has the discrete set of true bound states with eigenvalues
$\lambda_n^{(b)}$. Eq. (\ref{linear}) with $V$ replaced by $V_{b}$ has been solved
numerically
and some of its eigenstates are shown in Fig.~\ref{fig5} with dashed lines.
Then we take the finite potential $V$ and for each $n$ find few eigenvalues
and corresponding eigenstates
in the spectral proximity of
$\lambda_n^{(b)}$, applying the zero boundary conditions at both ends.
The latter implies that we reliably calculate only the states
with relatively small tail amplitude at $\xi\to -\infty$. Having attempting
more precise calculations would go beyond our original level of precision anyway,
because the oscillations of the soliton tail at $\xi\to -\infty$
have been disregarded in the first place. Some quasi-bound eigenstates of
the true potential $V$ are shown in Fig.~\ref{fig5}(b)-(e) with the full lines,
and the corresponding eigenvalues are indicated in Fig.~\ref{fig5}(a).

The terms $\exp\left[\xi \dot{x}_0/(2d_{1,2})\right]$ in Eqs. (\ref{transform_sol}), (\ref{transform_rad})
explicitly express the continuous frequency shifts of the two components
with the rates $g/2d_{1,2}$. The anomalous GVD ($d_1>0$)
corresponds to the  expected red frequency shift \cite{Gordon1986}. However,
the normal GVD ($d_2<0$) implies the blue frequency shift. The latter explains
spectral dynamics of the trapped radiation. Note, that some analogy may exist
here with the previously reported blue shift of the spectral holes associated with the
dark fiber solitons existing in the normal GVD range \cite{tomlin}.

Equations (\ref{transform_sol}), (\ref{transform_rad}) give, however, little physical
insight into which elementary wave scattering mechanisms lead to these opposing frequency shifts.
The physical process driving the red
shift of the soliton component is the well known intrapulse Raman scattering \cite{agrawal,Gordon1986}.
The blue shift of the radiation component
is driven by the intrapulse four-wave mixing described in details in \cite{GSS+2006}.
Briefly, it means that the scattering of the radiation pulse on the soliton  generates
the blue shifted pulse. The continuous frequency shift of the soliton
and the phase matching conditions work out in such a way
that the  observable result of this process
is the continuous blue shift of the radiation pulse carrier frequency. Trapping effect
sustains this process over the long propagation distances and results in existence of the
stationary soliton-radiation states.

In order to verify validity of our approximate eigenvalue analysis
we have initialized Eqs. (\ref{CNLS1}), (\ref{CNLS2})
with the soliton for $A_1$ and the linear eigenstates of $V$
for $A_2$ and solved the equations numerically. Fig.~\ref{fig6} shows evolution
of the first three quasi-bound eigenstates. The amplitude of the
eigenstates has been kept small in order to ensure that we remain
in the regime, when $|A_2|^2$ terms are negligible. One can see that, despite
using the simplified boundary conditions, our solutions satisfy well
the coupled NLS equations. This simulation also confirms that
the propagation distances on which the tunneling induced losses
lead to noticeable effects are  much larger than the typical GVD length.
So that for the fiber length of order meters the dispersive spreading
of the high frequency radiation is suppressed and it propagates as a localized
state of light. Note also, that the characteristic
feature of each quasi-bound state, apart from the
lowest one, is the presence of several spectral peaks.
Such multi-peak spectral structures are typical for the blue wing
of supercontinua seen in Figs.~\ref{fig2}(a), (c).

\begin{figure}
\includegraphics[width=0.48\textwidth]{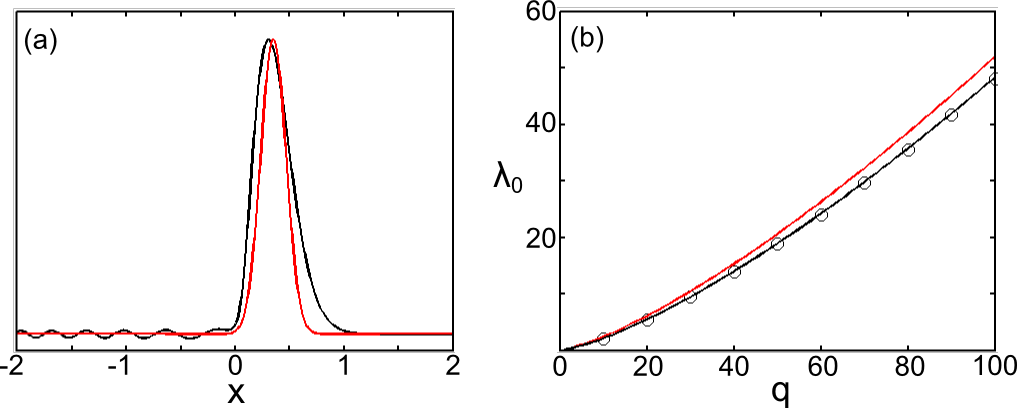}
\caption{
(Color online)
(a) Lowest trapped linear mode of the potential $V$ (solid line)
and its variational approximation (dashed line), given by Eq. (\ref{var_ansatz}) with
$\gamma$ and $w^2$ defined through (\ref{var_solution_g}) and (\ref{var_solution_w}),
respectively. Soliton parameter is $q=100$.
(b) $\lambda_0^{(b)}(q)$, solid line, $\lambda_0(q)$, circles,
and $\tilde\lambda_0(q)$, dashed line.}
\label{fig_var_mode}
\end{figure}

Approximation $V\approx V_{b}$ is also useful because it
allows us to carry out explicit variational calculations of the
eigenvalues and eigenfunctions, and thus to have  analytical
estimates for the width of the trapped states. Let us consider the
variational approximation for the ground state ($n=0$) only.
As a trial function we choose
\begin{equation}
\label{var_ansatz}
\tilde\phi_0=\exp\left[-(\xi-\gamma)^2/w^2\right]\;,
\end{equation}
where $\gamma$ is the shift of the intensity maximum of the trapped state
with respect to the soliton one and $w$ is the width of trapped state.
One can suggest a better trial functions accounting for the asymmetry of the
profile of the ground. However, our choice is the best suited for getting
transparent analytical expressions for parameters $\gamma$ and $w$.
The variational estimate
$\tilde\lambda_0$ for the true value $\lambda_0\le \tilde\lambda_0$ is
\begin{equation}
\label{var}
\tilde\lambda_0=\frac{\int d\xi \;\tilde\phi_0  \left[-|d_2| \partial^2_\xi +
V_{b}  \right]\tilde\phi_0}{\int d\xi \; \tilde\phi_0^2}
\end{equation}
Minimizing $\tilde\lambda_0$ with respect to $\gamma$ and $w^2$ we
find
\begin{eqnarray}
\label{var_solution_w}
w^2&=&2\sqrt{\frac{2|d_2|}{BC}}=\frac{|d_2|}{2q}\sqrt{\sqrt{\frac{d_1}{q}}\frac{15}{2T}},\\
\label{var_solution_g}
\gamma&=&\frac{B}{4}\sqrt{\frac{2|d_2|}{BC}}-\frac{1}{B}\ln\left(\frac{C}{BA}\right),\\
\label{var_solution_lam}
\tilde\lambda_0&=&\sqrt{\frac{|d_2|BC}{2}}+\frac{C}{B}\left[1-\ln\left(\frac{C}{BA}.
\right)\right]
\end{eqnarray}
Thus the narrow solitons (large $q$) and large Raman effect (large $T$) result,
quite naturally, in stronger localization of the radiation.
Figure~\ref{fig_var_mode}(a) illustrates comparison between the numerically calculated
linear mode $\phi_0$ of the potential $V(\xi)$ (\ref{linear_pot})
and the variational approximation.
The numerically calculated dependencies $\lambda_0(q)$ and $\lambda_0^{(b)}(q)$ for
potentials $V$ and $V_b$, respectively, compare well with the variational result (\ref{var_solution_lam}),
see Fig.~\ref{fig_var_mode}(b).

\section{Nonlinear theory of the two-frequency quasi-solitons across the zero GVD point}

The theory presented above explains  the nature of the radiation
trapping and is adequate in the regime when the radiation is weak.
In this regime the family of the soliton-radiation bound states is
continuously parameterized by $q$ only, i.e. by the amplitude of the
soliton pulse. In particular, this is expressed in the fact that the
acceleration parameter $g=g_0$ and the eigenvalue $\lambda$ are
fixed for a given $q$. In the nonlinear regime one should expect
that $\lambda$ will become a continuously varying parameter, as it
happens for other types of localized solutions in incoherently
coupled  NLS equations \cite{BKP1996}. Thus  the acceleration
$g$ is expected to be  continuously parameterized by both
$q$ and $\lambda$, i.e. by the energies of the both fields
$\phi$ and $\psi$. In order to demonstrate explicitly, that indeed there is a
problem to be addressed here, we compare the propagation distance at
which the spectral recoil from the Cherenkov radiation stabilizing
the soliton frequency takes place with and without the blue detuned
pulse seeded into the fiber. One can see, Figs.~\ref{fig4}(a) and
(e), that the sharp transition to the regime without the
self-frequency shift happens at a shorter distance with the blue radiation present.
This is because the spectral peak of the soliton reaches the
critical distance from the zero GVD point sooner, which indicates
that the rate of the self-frequency shift is faster for  the soliton-radiation
bound state than for the pure soliton, i.e.  $g>g_0$. Thus
nonlinear in $\phi$ corrections should be taken into account to
explain this effect.

One obvious  small parameter in our problem is the  Raman time $T$, which enters both
equations for $\phi$ and $\psi$.
The linear in $\xi$ potential
can be considered as a small perturbation only in the equation for $\psi$. However,
it plays a crucial role in the localization of the
$\phi$ component at $\xi\to\infty$ and therefore can not
be neglected already in the leading order in the equation for $\phi$.
As we have found above, see Figs.~\ref{fig5} and
\ref{fig6}, the $\phi$ and $\psi$ components overlap only by their tails.
Therefore the nonlinear coupling can be considered as a small perturbation on the $\psi$ component,
which is self-localized.
However, in the $\phi$ equation the term $|\psi|^2\phi$ is the only localization
mechanism for $\xi\to-\infty$ and hence can not be neglected there.
Now we rewrite Eqs.~(\ref{CNLS1_w}),
(\ref{CNLS2_w})
collecting all the leading  terms in the left hand-side, all the first order corrections
in the right-hand side and neglecting the rest:
\begin{eqnarray}
\nonumber && i\partial_z \psi + d_1 \partial^2_\xi \psi - q \psi +
|\psi|^2\psi\\&& =-2|\phi|^2\psi+T \psi \partial_\xi |\psi|^2 + \frac{g\xi}{2d_1}\psi+\dots,
\label{CNLS1_n}\\
\nonumber && i\partial_z \phi - |d_2| \partial^2_\xi \phi -
\lambda \phi +|\phi|^2\phi+2|\psi|^2\phi+ \frac{g\xi}{2d_2}\phi\\&&=
T \phi \partial_\xi|\phi|^2 +\dots
\label{CNLS2_n}
\end{eqnarray}
In order to calculate the deviation of $g$ from $g_0$, we  assume
that $\psi=\psi_{0}(\xi)+\psi_{1}(\xi)+\dots$,
$\phi=\phi_{0}(\xi)+\phi_{1}(\xi)+\dots$, where $\psi_1$ and $\phi_1$
have the same order of smallness as the right-hand sides in Eqs. (\ref{CNLS1_n}), (\ref{CNLS2_n}).
This leads to
\begin{eqnarray}
\nonumber && d_1\partial_{\xi}^2\psi_1-q\psi_1+3\psi_0^2\psi_1\\
&& =T\psi_0\partial_{\xi}\psi_0^2+{g\xi\over
2d_1}\psi_0-2\phi_{0}^2\psi_0,
\label{psi11}
\end{eqnarray}
where $\phi_0$ obeys
\begin{eqnarray}
-|d_2| \partial^2_\xi \phi_0 +V(\xi,g)\phi_0=\lambda\phi_0-|\phi_0|^2\phi_0  \label{nonlinear}
\end{eqnarray}

\begin{figure}
\includegraphics[width=0.48\textwidth]{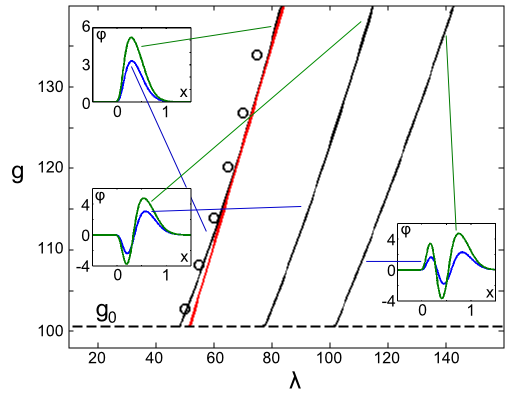}
\caption{
(Color online)
Numerically calculated acceleration $g$ for different branches
of solutions of Eqs.~(\ref{nonlinear}), (\ref{gcor}) with the potential $V$
being replaced by its asymptotic $V_b$.
Circles correspond to the solutions in the full potential $V$. Approximate analytical solution
(\ref{dg_approx}) is indicated by dashed line. Insets illustrate solution profiles (asymptotic potential)
at different values of the parameter $\lambda$.}
\label{fig8}
\end{figure}

Projecting the right-hand side of Eq. (\ref{psi11}) on
$\partial_{\xi}\psi_0$. We find
\begin{equation}
g=g_0+2\sqrt{d_1\over q}\int_{-\infty}^{\infty}d\xi
\phi_{0}^2\psi_0\partial_{\xi}\psi_0, \label{gcor}
\end{equation}
One should remember though that $\phi_{0}$ itself is a function of $g$.
Therefore Eq. (\ref{gcor}) is an equation for $g$, which needs to be solved.
Solving Eq. (\ref{nonlinear}) numerically
we find the set of functions $\phi_{0}$ parameterized by $g$ and $\lambda$,
which are the direct continuations of the linear discrete set of eigenfunctions of the potential
$V$ found in the previous section.
The tail of the solution for $\xi\to-\infty$ is oscillatory and weakly decaying
like the tail of the linear solutions.
However, despite the fact that $\phi_{0}$ is  only semi-bound, it can be used
in the integral (\ref{gcor}), because it is multiplied there by the exponentially localized
functions $\psi_0$ and $\partial_\xi \psi_0$.
It reflects the fact that the tail of $\phi$ has only a negligible contribution into
the selection of $g$.
Therefore, like in the previous section, we
can replace $V$ with $V_b$ and carry out calculations using the
infinite potential well and exponentially decaying solutions.
Substituting $\phi_{0}$ inside the condition (\ref{gcor})
and solving the latter numerically for $g$, we find the corrected values of $g$.
Fig.~\ref{fig8} shows dependencies of $g$ on $\lambda$ for the first three
bound states. One can see that  increase in the  amplitude of the $\phi$-component
generally leads to the larger values of $g$, which
practically means stronger negative accelerations and larger
frequency shifts (the red/blue shift for the $\psi$/$\phi$-component, respectively).

It is useful to derive an approximate analytical expression for $g$, which can be done
in several ways. First, the variational approach can be applied to the nonlinear problem
(\ref{nonlinear}). However, this leads to the rather cumbersome  and difficult to
understand expressions. More elegant answer, which also matches well our numerical
calculations in Fig.~\ref{fig8},
can be obtained in the limit of the weak nonlinearity
in Eq. (\ref{nonlinear}). This is accomplished by solving Eq. (\ref{nonlinear})
perturbatively. We consider the ground state and assume that in the first order
$\phi_0$ and $\lambda$ are  given by their
variational approximations $A\tilde\phi_0$ and $\tilde\lambda_0$
found in the linear case, see Eqs.~(\ref{var_ansatz}), (\ref{var}).
Here $A$ is the constant amplitude to be determined.
We also introduce  correction to the eigenfunction,
$\delta\phi$,  and to the eigenvalue,
$\delta\lambda=\lambda-\tilde\lambda_0$,
 induced by the nonlinearity. Taking $\epsilon$ as a dummy small parameter,
 we  assume  $A^2\sim \epsilon$, $\delta\phi\sim\epsilon$ and $\delta\lambda\sim\epsilon$. Looking
 at the integral in Eq. (\ref{gcor}) we see that it has order $\epsilon^2$,
 because its value is proportional to $A^2$ and it acquires an extra order of smallness
 due to a small overlap between $\phi_0$ and $\psi_0$.
 It means that $g=g_0+O(\epsilon^2)$.
 The  resulting equation for $\delta\phi$ derived from Eq. (\ref{nonlinear}) is
\begin{eqnarray}
\nonumber
(-|d_2| \partial^2_\xi  +V(\xi,g_0)-\tilde\lambda_0)\delta\phi\approx\\
\delta\lambda A\tilde\phi_0
-A^3\tilde\phi_0^3.
\label{linear_cor}
\end{eqnarray}
The operator in the left hand side is self-adjoint and singular. Its null-space
is given by the linear ground state. Therefore projecting the right-hand side on $\tilde\phi_0$
we find $\delta\lambda=\sqrt{2}A^2$.
Substituting $\phi_0$ in Eq.~(\ref{gcor})
with $A\tilde\phi_0$
and taking into account that inside the integral in Eq.~(\ref{gcor})
the $\psi_0$ can be approximately replaced with
its asymptotic  $2\sqrt{2q}\exp[-\xi\sqrt{q/d_1}]$
 we find a relatively simple expression for $g$:
\begin{equation}
\label{dg_approx}
g \approx\tilde{g}= g_0 \cdot \left[
1+\frac{\lambda-\tilde{\lambda}_0}{q^{9/8}T^{1/4}}\frac{d_1^{5/8}}{2|d_2|} \sqrt{\frac{\pi}{2}}
\left(\frac{15|d_2|}{32}\right)^{1/4}
\right].
\end{equation}
One can see that the above assumption about smallness of the overlap
between $\phi_0$ and $\psi_0$ remains valid provided
$q^{-9/8}T^{-1/4}\sim\epsilon$. With $T^{1/4}$ being practically order of $1$,
it implies that $q$ should be sufficiently large.
Together with the variational approximation
for $\tilde{\lambda}_0$, see Eq.~(\ref{var_solution_lam}), the above
equation agrees very well with numerically calculated dependence of
$g$ on $\lambda$ for the ground states, see Fig.~\ref{fig8}.

\begin{figure}
\includegraphics[width=0.48\textwidth]{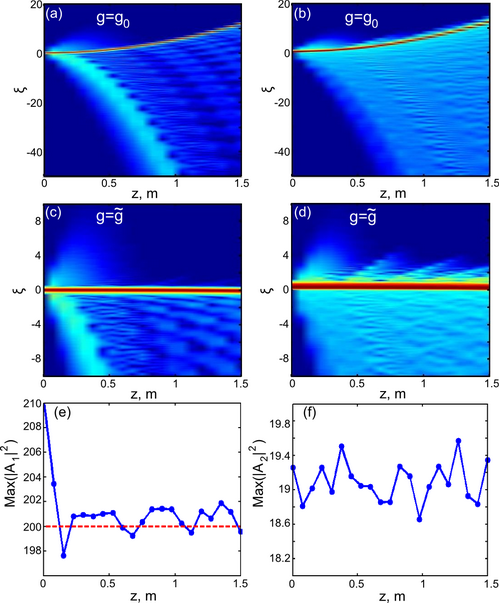}
\caption{
(Color online)
Numerical propagation of the lowest bound state within coupled equations
(\ref{CNLS1_w}) and (\ref{CNLS2_w}) with acceleration $g=g_0$, (a) and (b), and $g=\tilde{g}$,
(c) and (d). Left column: anomalous GVD component ($A_1$), right column: normal GVD component ($A_2$).
Intensities are plotted in logarithmic scale.
(e) and (f): maximum of the intensity of $A_1$ and $A_2$ components, respectively,
along propagation distance.
Initial shape for $A_2$ was calculated with the approximate potential $V_b$ for
the soliton parameter $q=100$ and $\lambda=70$
($\delta\lambda\approx 20$). $A_1$ was initialized with the soliton (\ref{soliton})
with $q=105$ in order to account for initial radiative losses, see text for more details.
}
\label{fig9}
\end{figure}

An ultimate method to confirm the validity of our approximate calculations is to take  Eqs.
(\ref{CNLS1_w}), (\ref{CNLS2_w}) in the reference frame moving with
acceleration $g=g_0$ and with $g=\tilde{g}$ given by Eq. (\ref{dg_approx}), initialize them with
the bound state and solve numerically. Then trajectories of the solutions
on the $(z,\xi)$-plane should be straight lines in the case of the exactly selected
acceleration and parabolic otherwise.
Fig.~\ref{fig9} demonstrates the results of this numerical experiment, where
intensities of the two components $A_1$ and $A_2$ are plotted separately.
Since initial excitation was taken in a localized form, i.e. without proper account
of oscillating tails, one can see that both components emit radiation during propagation.
The radiation leakage from the trapped  component, $A_2$, is small
[note logarithmic scale in Figs.~\ref{fig9} (a)-(d)] and does
not considerably affect the acceleration of the bound state.
More important, however, is the initial outburst of radiation from the soliton
component, $A_1$. As the result, a noticeable drop in intensity
of $A_1$ ($\sim 5\%$) is observed during initial stage of soliton propagation,
see Fig.~\ref{fig9} (e), while the intensity of the $A_2$ component stays practically the same
during propagation, see Fig.~\ref{fig9} (f).
In order to account for these losses,
in the initial condition for $A_1$ we have increased
the soliton parameter $q$ by $5\%$ with respect
to the value used in the calculations of the acceleration $\tilde{g}$.
Taking this into account,
the  Eq.~(\ref{dg_approx}) gives a very good approximation for the
acceleration of the bound state and, hence, for the rate of the self-frequency shift
associated with it, see Figs.~\ref{fig9}(c) and (d).

\section{Summary}
In summary, we have presented the detailed theory of the  effect of radiation trapping
by the Raman accelerated fiber solitons responsible for formation of the blue
wing of the supercontinuum spectra.
We demonstrated, that the radiation in
the range of the normal GVD is subject to the  inertial force, which,
together with the soliton induced refractive index change,
forms an effective potential well prohibiting  the
dispersive spreading of the radiation.
We have found not only the ground state of the radiation field, but also its excited states,
which  relevance for the past and ongoing experimental observations
is under current investigation.
The soliton-radiation bound states move with a constant acceleration in the time-space,
while in the spectral domain the peaks corresponding to the two components
move in opposite spectral directions (soliton component always gets redder, while
the radiation component gets bluer).

In the first part of our theoretical considerations  we have assumed that the radiation is linear.
In this case we have demonstrated that the continuous
blue  shift of its frequency happens at
the same rate as has been previously calculated for the red shifting solitons \cite{agrawal,Gordon1986}.
Considering effects nonlinear in the radiation,
we have found that the acceleration of the soliton-radiation
bound states increases with the radiation amplitude, which corresponds  to
larger rates of the self-frequency shift. We have also derived
an approximate analytical expression for the latter.

The results presented above pave the way for design of new fiber based
soliton frequency converters, which allow for efficient blue frequency shifts
of the soliton like state. This removes traditional restriction of
the Raman  soliton based frequency conversion
been directed only towards longer wavelengths.
Our results also emphasize  that moving dielectric media
can be created via nonlinear modulation of the refractive index by the  pulse
propagating with a speed of light. This creates an interesting testing bed for
studies of the effects of light propagation in moving dielectrics,
which have  generated significant recent interest, see e.g. \cite{leon}.

\begin{acknowledgments}
This work has been supported by EPSRC.
\end{acknowledgments}


\end{document}